\documentclass[aps,prb,preprint,superscriptaddress]{revtex4-1}

\usepackage{graphicx}
\usepackage{dcolumn}
\usepackage{bm}
\usepackage[breaklinks]{hyperref}
\usepackage[all]{hypcap}
\usepackage{color}
\usepackage{array}

\newcommand{\e}{\mathrm{e}}

\begin{document}


\title{Do Fourier analysis yield reliable amplitude of quantum oscillations?}

\author{Alain Audouard}
\affiliation{Laboratoire National des Champs Magn\'{e}tiques
Intenses (UPR 3228 CNRS, INSA, UGA, UPS) 143 avenue de Rangueil,
F-31400 Toulouse, France.}

\author{Jean-Yves~Fortin}
\affiliation{Institut Jean Lamour, D\'epartement de Physique de la
Mati\`ere et des Mat\'eriaux,
CNRS-UMR 7198, Vandoeuvre-l\`{e}s-Nancy, F-54506, France.
}%

\date{\today}


%

\begin{abstract}

Quantum oscillations amplitude of multiband metals, such as high T$_c$ superconductors in the normal state, heavy fermions or organic conductors are generally determined through Fourier analysis of the data even though the oscillatory part of the signal is field-dependent. It is demonstrated that the amplitude of a given Fourier component can strongly depend on both the nature of the windowing (either flat, Hahn or Blackman window) and, since oscillations are obtained within finite field range, the window width. Consequences on the determination of the Fourier amplitude, hence on the effective mass are examined in order to determine the conditions for reliable data analysis.

\end{abstract}


\maketitle

\section{Introduction}

Quantum oscillations, the extrema of which are periodic in inverse magnetic field, are known to provide valuable information for the study of Fermi surface of metals. In particular, in addition to their frequency which yields Fermi surface cross section, field and temperature dependence of their amplitude allows for
determination of the effective mass and scattering rate \cite{Sh84}.
Multiband metals such as heavy fermions \cite{On12} or high-T$_c$ superconducting iron chalcogenides \cite{Te14,W15,Wa15,Au15f} have complex Fermi surface due to numerous sheets crossing the Fermi level, giving rise to  many orbits in magnetic field, hence to complex quantum oscillation spectra. Besides, in the case where magnetic breakdown (MB) between orbits occurs, as it is the case of many organic metals \cite{Uj08,Au13}, additional orbits are further generated. In such cases, data can be readily derived through Fourier analysis, allowing discrimination between the various frequencies.
The point is that the amplitude of quantum oscillations is field-dependent. Therefore, strictly speaking, they are not periodic in inverse field. More specifically, at a fixed temperature $T$, a given Fourier component of the oscillatory part of magnetization (de Haas-van Alphen oscillations) and conductivity (Shubnikov-de Haas oscillations) can be written as $A(x)=A_0(x)\sin(2\pi f_0x+ \phi)$ where $x=1/B$, $f_0$ is the frequency and $\phi$ is, for normal metals, the Onsager phase. In the framework of the Lifshitz-Kosevich and Falicov-Stachowiak models \cite{Sh84}, the amplitude is given by $A_0(x) \propto R_T R_D R_{MB}$ for a given field direction (in which case the spin damping factor is a field- and temperature-independent prefactor). For a two-dimensional orbit, the thermal, Dingle and MB damping factors are given by  $R_T=u_0Tm^*x/\sinh(u_0Tm^*x)$, $R_D = \e^{-u_0T_Dm^*x}$ and $R_{MB} = \e^{-n_tB_0x/2}[1-\e^{-B_0x}]^{n_r/2}$, respectively, where $u_0$ = 2$\pi^2 k_B m_e(e\hbar)^{-1}$ = 14.694 T/K, $m^*$ is the effective mass and $T_D$ is the Dingle temperature, ($T_D$= $\hbar/2\pi k_B\tau$, where $\tau$ is the relaxation time). $n_t$ and $n_r$ are the number of tunneling and reflections the quasiparticles are facing during their travel along a MB orbit with a MB gap $B_0$.  The question that arises is then to determine to what extent reliable oscillation amplitudes can be derived from Fourier analysis of such field-dependent data.

\begin{figure}
\resizebox{0.75\columnwidth}{!}{%
  \includegraphics{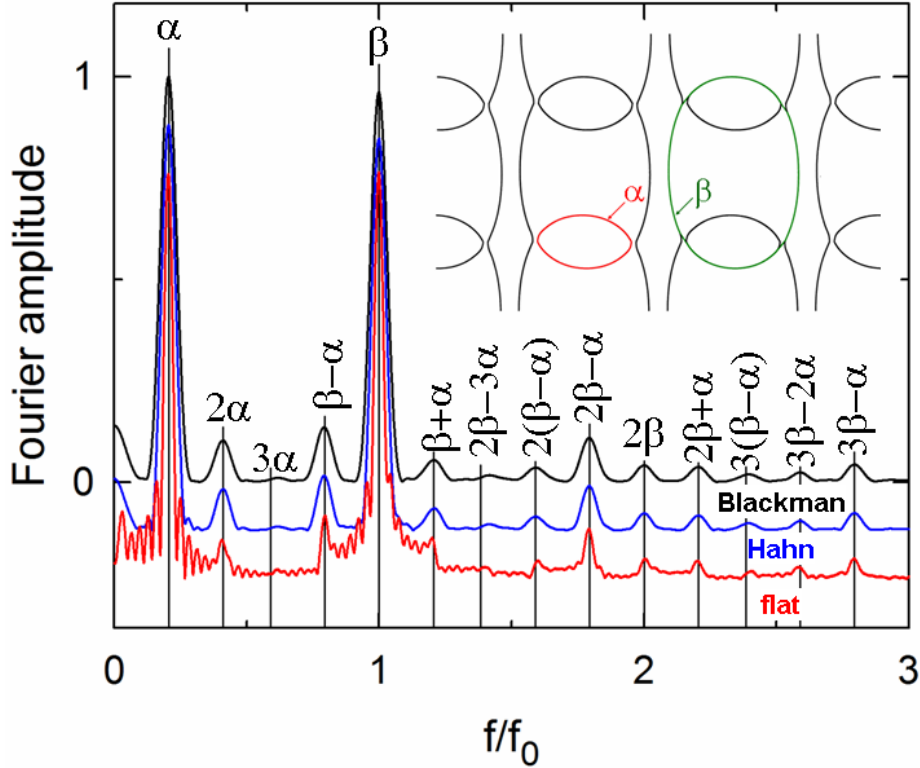}
}
\caption{Fourier analysis of de Haas-van Alphen oscillations of the organic metal
$\theta$-(ET)$_4$ZnBr$_4$(C$_6$H$_4$Cl$_2$), obtained with Blackman, Hahn and flat windows in the field range 40-56 T at 2K. Vertical lines are marks calculated with $f_{\alpha}$ = 930 T and $f_0=f_{\beta}$ = 4534 T.
The insert displays the Fermi surface in which the $\alpha$ and $\beta$ orbits are indicated (data are from Ref.~\cite{Au15}). }
\label{Fig_ET4Zn}       
\end{figure}

In the following, the organic metal $\theta$-(ET)$_4$ZnBr$_4$(C$_6$H$_4$Cl$_2$), the de Haas-van Alphen and Shubnikov-de Haas oscillations of which were extensively studied in pulsed magnetic fields of up to 55 T \cite{Au15} (see Fig.~\ref{Fig_ET4Zn}), is considered. As it is the case of many compounds based on the bis(ethylenedithio)tetrathiafulvalene molecule (abbreviated as ET), this compound illustrates the model Fermi surface proposed by Pippard to compute magnetic breakdown amplitudes of multiband metals \cite{Pi62}. As reported in Fig.~\ref{Fig_ET4Zn}, its Fermi surface is composed of one strongly two-dimensional closed orbit ($\alpha$) and a pair of quasi-one dimensional sheets giving rise in magnetic field to the MB orbit $\beta$. As a result, oscillation spectra are composed of many frequencies which are linear combinations of the frequencies linked to the $\alpha$ and $\beta$ orbits. Amplitudes relevant to these combinations are strongly influenced by oscillations of the chemical potential in magnetic field \cite{Au12,Au15}. Nevertheless, this phenomenon has negligible influence on the amplitude of the basic components $\alpha$ and $\beta$ allowing relevant data analysis on the basis of the above mentioned Lifshitz-Kosevich formalism.

Rather than bringing additional information on this compound, the aim of this paper is to determine to what extent Fourier analysis is able to yield reliable values of physical parameters of interest, in particular effective mass and scattering rate (through the Dingle temperature). To this purpose, we will consider the  $\beta$ orbit, with frequency $f_0$ = $f_{\beta}$ = 4534 T, effective mass $m^* = m_{\beta}$ = 3.4  $m_e$ and $T_D$ = 0.8 K (this latter parameter being dependent on the considered crystal), which involves no reflections ($n_r$ = 0) and 4 tunnelings ($n_t$ = 4) with MB field $B_0$ = 26 T \cite{Au15}. This component will serve as a basis to determine the influence of the windowing (nature and width) on the Fourier amplitude evaluation.

\section{Methodology}
\label{sec:methodology}

In the following we will consider dHvA oscillations relevant to the above mentioned $\beta$ orbit. Since measured magnetic torque $\boldsymbol{\tau}$ is related to magnetization
$\mathbf{M}$  as $\boldsymbol{\tau} = \mathbf{M \times B}$, Fourier amplitude can be written:
\begin{eqnarray}
\label{Eq:LK}
 A_0(x) \propto \frac{Tm_{\beta}}{\sinh(u_0Tm_{\beta}x)}\exp[-(u_0T_Dm_{\beta}+2B_0)x]
\end{eqnarray}
At high enough values of $u_0Tm^*x$, $A_0(x)$ can be approximated as
\begin{eqnarray}
\label{Eq:lambda}
 A_0(x) \simeq a_0\exp(-\lambda x)
\end{eqnarray}
where $a_0$ is a temperature-dependent prefactor ($a_0 \propto T$) and $\lambda = u_0(T+T_D)m_{\beta}+2B_0$\cite{footnote1}. This approximation provides a single parameter characterizing the field dependence of the amplitude: the largest $\lambda$, the steepest the field dependence. For $\theta$-(ET)$_4$ZnBr$_4$(C$_6$H$_4$Cl$_2$), explored $\lambda$ values are  within 194 T at 2 K and 305 T at 4.2 K. Due to large Dingle temperature, even larger values are obtained for the high-T$_c$ superconductor FeSe for which $\lambda$ varies from 250 T at 1.6 K to 370 T at 4.2 K \cite{Au15f}.

Since the signal amplitude is field-dependent, windowing \cite{Fi64,Al77,Ha78,Ac90,Na07} is mandatory in order to determine Fourier amplitude at a given inverse field value $ \bar x$. The inverse field range $\Delta x$ is within $x_m$ and $x_M$  ($\Delta x = x_M-x_m$) and centered on $ \bar x=(x_m+x_M)/2$. In order to explore the influence of windowing on the Fourier amplitude, flat, Hahn and Blackman windows are considered in the following: $w(x) = 1$, $w(x) = \{1 + \cos[2\pi(x- \bar x)/\Delta x]\}/2$ and $w(x) = 0.42 + 0.5\cos[2\pi(x- \bar x)/\Delta x] + 0.08\cos[4\pi(x- \bar x)/\Delta x]$, respectively, within the range $x_m$ to $x_M$ and $w(x)$ = 0 everywhere else. We can write more generally the window function as $w(x)=\sum_{n\ge 0}^pc_n\cos[2\pi n(x-\bar x)/\Delta x]$, where $p=0,1,2$ for a flat, Hahn and Blackman window respectively, but can be generalized for higher values of $p$. Discrete Fourier transforms are obtained as

\begin{eqnarray}
\label{Eq:F(f)}
F(f,\bar x)=\frac{2}{\Delta x}\int_{x_m}^{x_M}A_0(x)\sin(2\pi f_0x+ \phi)w(x)\exp(-2i\pi f x)dx,
\end{eqnarray}

Analytical solution of Eq.~\ref{Eq:F(f)} is given in the Appendix (Eq.~\ref{Eq:Ff0}) for $f = f_0$. Modulus of $F(f_0,\bar x)$ yields the Fourier amplitude $A_F(\bar x)=|F(f_0, \bar x)|/c_0$. For finite $\lambda$ and $f_0\gg\lambda$, Eq.~\ref{Eq:F} holds, yielding

\begin{eqnarray}
\label{Eq:Fmain}
A_F(\bar x)=A_0(\bar x)c_0^{-1}
\frac{\sinh(\lambda\Delta x/2)}{\lambda \Delta x/2}
\sum_{n\ge 0}(-1)^nc_n
\frac{(\lambda \Delta x/2)^2}{
(\lambda \Delta x/2)^2+\pi^2n^2}
\end{eqnarray}

$A_F(\bar x)$ can also be obtained by numerical resolution of Eq.~\ref{Eq:F(f)} where $A_0(\bar x)$ is either given by Eq.~\ref{Eq:lambda} or by experimental data of Ref.~\cite{Au15}.
Available frequencies are bounded by the Raleigh frequency ($f_{min}=1/\Delta x$) and by the Nyquist frequency ($f_{max}$ = 1/2$\delta x$, for data sampled at evenly spaced $\delta x$ values). Accordingly, $\Delta x$ is kept above $1/f_0$ and $\delta x$ is always small enough to ensure that $f_{max}$ is much higher than $f_0$ \cite{Dr10} in the following.


\section{Results and discussion}

\begin{figure}
\resizebox{0.9\columnwidth}{!}{%
  \includegraphics{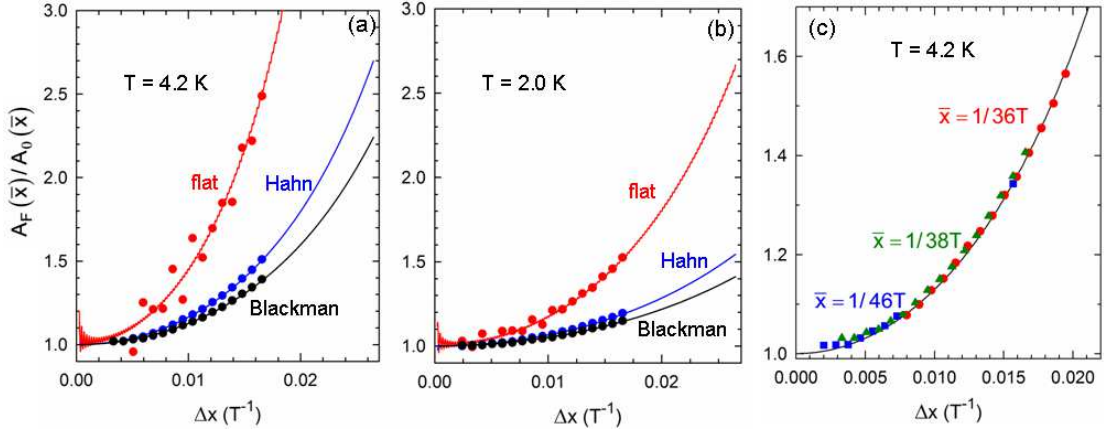}
}
\caption{Fourier amplitude $A_F(\bar x)$ relevant to the $\beta$ component of the organic metal $\theta$-(ET)$_4$ZnBr$_4$(C$_6$H$_4$Cl$_2$) at (a) 4.2 K ($\lambda$ = 305 T)
and (b) 2.0 K ($\lambda$ = 194 T), normalized to the oscillation amplitude predicted by the Lifshitz-Kosevich formula $A_0(\bar x)$ as a function of the inverse field window
$\Delta x$ for flat, Hahn and Blackman windows, at $ \bar x$ = 1/38 T and (c) for Blackman window at various $ \bar x$ values. Solid symbols are deduced from
experimental data reported in Ref.~\cite{Au15}. }
\label{Fig_ET4Zn_fenetre}       
\end{figure}

\begin{figure}
\resizebox{0.9\columnwidth}{!}{%
  \includegraphics{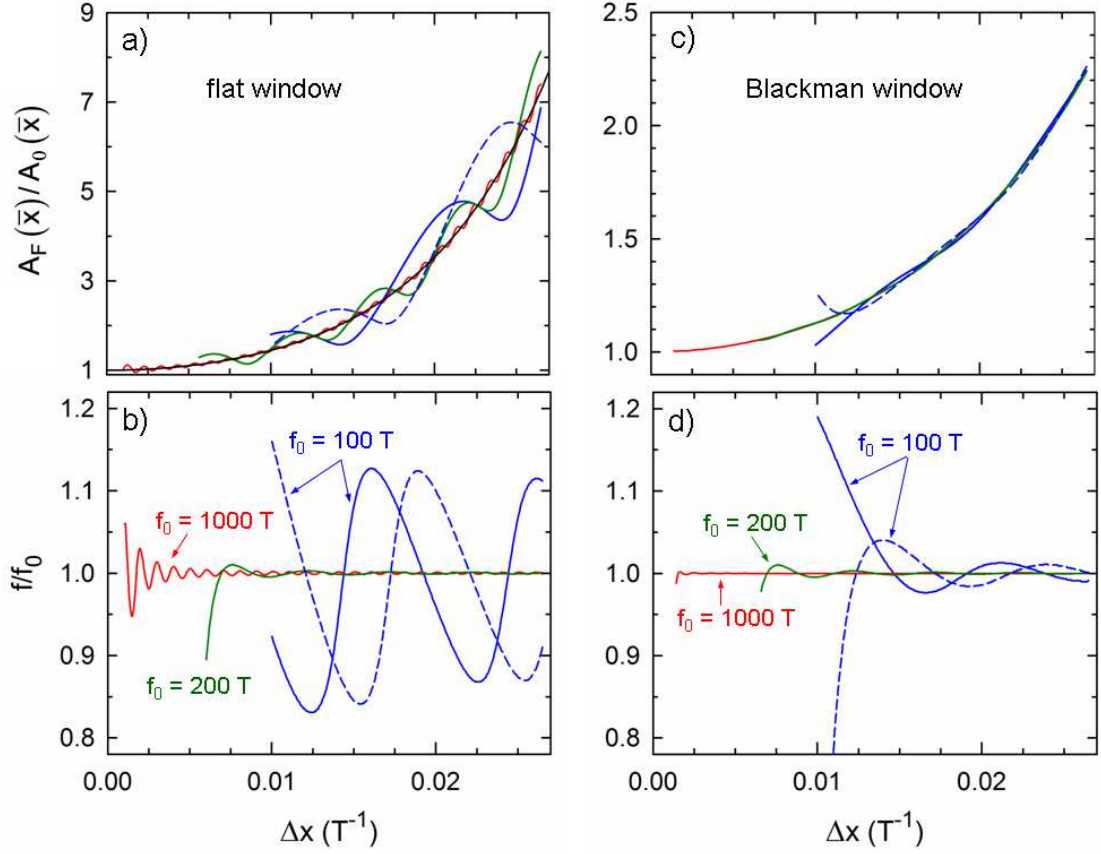}
}
\caption{Inverse field window width ($\Delta x$) dependence of (a), (c) Fourier amplitude and (b), (d) frequency for (a), (b) flat and (c), (d) Blackman window
for various oscillation frequencies $f_0$ and $\lambda$ = 305 T. The Onsager phase is $\phi$ = 0 and $\phi= \pi$ for solid and dashed lines, respectively. Black solid line in
(a) stands for Eq.~\ref{Eq:Fmain}.}
\label{Fig_f0}       
\end{figure}

\begin{figure}
\resizebox{0.9\columnwidth}{!}{%
  \includegraphics{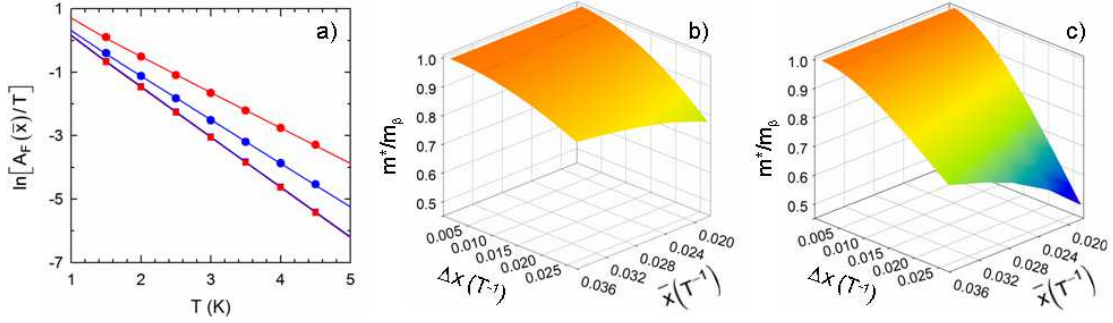}
}
\caption{(a) Mass plots relevant to the $\beta$ component of the organic metal $\theta$-(ET)$_4$ZnBr$_4$(C$_6$H$_4$Cl$_2$), with effective mass $m_{\beta}$ = 3.44, deduced from Fourier analysis for $ \bar x$ = 1/32 T,
in the temperature range 1.5K - 4.5 K. Blue and red symbols are data obtained with Blackman and flat windows, respectively. Solid squares and circles are data for $\Delta x$ = 0.00093 $T^{-1}$ and 0.0265 $T^{-1}$, respectively. Solid lines are best fits to the Lifshitz-Kosevich formula. For large $\Delta x$, both the Fourier amplitude increases and the slope decreases yielding underestimated effective mass.   Such fittings yield data of Figs.~\ref{Fig_ET4Zn_masseeffective}(b) and (c) where effective mass is plotted vs. inverse field window width ($\Delta x$) and mean inverse field value ($ \bar x$) for (b) Blackman and (c) flat window. At
high field (i.e. small $ \bar x$) and large field window width ($\Delta x$), strong underestimation of the effective mass is obtained.}
\label{Fig_ET4Zn_masseeffective}       
\end{figure}

Fourier analysis displayed in Fig.~\ref{Fig_ET4Zn} evidence that largest (smallest) secondary lobes and smallest (largest) peak width are obtained for the flat (Blackman) window while the Hahn window provides intermediate behaviour, as widely reported \cite{Fi64,Al77,Ha78,Ac90,Na07}.

Discrepancy between amplitude $A_F( \bar x)$ deduced from Fourier analysis within a finite field range $1/x_{max}$ to $1/x_{min}$ and the actual Fourier
amplitude $A_0( \bar x)$ given by Eq.~\ref{Eq:lambda} can be evaluated through the ratio $A_F( \bar x)/A_0( \bar x)$ which should be equal to 1.
According to the data in Fig.~\ref{Fig_ET4Zn_fenetre}, a strong increase of this ratio is observed as $\Delta x$ increases. Furthermore, for a given window
width $\Delta x$, it increases as $\lambda$ increases e.g. by increasing the temperature while, as the mean magnetic field ($1/ \bar x$) decreases, it grows
staying on the same curve, as reported in Fig.~\ref{Fig_ET4Zn_fenetre}(c). The most dramatic effect is observed for the flat window, indicating that smooth windowing
is necessary to get amplitudes as reliable as possible since, more specifically, $A_F( \bar x)/A_0( \bar x)$ grows as
$\sinh(\lambda \Delta x/2)/(\lambda \Delta x/2)$ in this case.

In line with Eq.~\ref{Eq:Fmain}, the ratio $A( \bar x)/A_0( \bar x)$ only depends on the product  $\lambda \Delta x$ for a given window type. Hence, strictly speaking, Fourier analysis yields reliable amplitude for finite $\Delta x$ in the case of field-independent signal ($\lambda$ = 0), only. Unfavorably, moderate oscillations of the Fourier amplitude are however observed for small $\Delta x$, in particular for the flat window. It can be checked that these oscillations are periodic in $\Delta x$, their frequency being just $f_0$, in agreement with Eq.~\ref{Eq:Ff0}. This feature brings us to consider the influence of the quantum oscillations frequency on the data. As reported in Fig.~\ref{Fig_f0}, Fourier amplitude $A_F( \bar x)$ is dominated by the monotonous term of Eq.~\ref{Eq:Ff0}, yielding Eqs.~\ref{Eq:F} and~\ref{Eq:Fmain}, in the case of large enough frequency and $\Delta x$. In contrast, large oscillations of both the Fourier amplitude and the frequency of the Fourier peaks (which is no more equal to $f_0$ in this case) are observed for low frequencies, which are relevant for $e.g.$ superconducting iron-based chalcogenides \cite{Au15f,W15}. In addition, whereas only the envelope of $A_F(\bar x)$, i.e. $A_0(\bar x)$, is relevant for the Fourier amplitude at high $\Delta x$, Onsager phase-dependent data are observed in Fig.~\ref{Fig_f0} for low frequencies. In short, $\Delta x$ must be both small enough to avoid the amplitude overestimation predicted by Eq.~\ref{Eq:Ff0} and large enough to avoid the undulations reported in Fig.~\ref{Fig_f0} in this case. As a consequence, reliable data can hardly been deduced from Fourier analysis of low frequency quantum oscillations.


Since $\lambda$ depends on temperature, the discrepancy between the actual and Fourier amplitudes for large $\Delta x$ depends on temperature as well. This may lead to significant error on the effective mass deduced from temperature dependence of the amplitude (so called mass plot), as evidenced in Fig.~\ref{Fig_ET4Zn_masseeffective}(a), hence on the determination of the scattering rate through Dingle plots, as well. As reported in Fig.~\ref{Fig_ET4Zn_masseeffective}(b) and (c), underestimation of $m_{\beta}$ by about 30 percent is obtained for a flat window at $ \bar x$ = 1/32 T$^{-1}$ for $\Delta x$ = 0.026 T$^{-1}$ (i.e. in the field range 23-56 T). About 50 percent would be reached  at $ \bar x$ = 1/56 T$^{-1}$ for the same $\Delta x$ value (field range within 32 and 193 T). Smaller although significant errors are obtained for Hahn (not shown) and Blackman windows, e.g.  15 and 13 percent, respectively, for  $ \bar x$ = 1/32 T$^{-1}$ and $\Delta x$ = 0.026 T$^{-1}$.

\section{Conclusion}

Amplitude of field-dependent quantum oscillations deduced from Fourier analysis is overestimated even though it is widely used, as reported in the literature. Most dramatic effects are observed for steep field-dependent amplitudes determined using flat windows with large width. Nevertheless, acceptable discrepancy with actual amplitude is obtained with Blackman window of moderate width for high enough frequencies. In contrast, oscillations with low frequencies such as observed in iron-based chalcogenides superconductors must be considered with care since $\Delta x$ must be both small enough to avoid overestimated amplitude and large enough to avoid spurious effects observed coming close to the inverse of the Raleigh frequency.

\acknowledgments
Work in Toulouse was supported by the European Magnetic Field Laboratory (EMFL). D. Vignolles, R.B. Lyubovskii, L. Drigo, G. V. Shilov, F. Duc, E. I. Zhilyaeva, R. N. Lyubovskaya and E. Canadell, as co-authors of Ref.~\cite{Au15} on which are based the data of Figs.~\ref{Fig_ET4Zn} and~\ref{Fig_ET4Zn_fenetre}, are acknowledged.

\appendix*
\section{Analytical expression of the Fourier transforms}

In general we can write the window function $w(x)=\sum_{n\ge 0}^pc_n\cos[2\pi n(x-\bar x)/\Delta x]$ where $p=0,1,2$ for a flat, Hahn and Blackman window respectively, and the condition $\sum_{n=0}^pc_n=1$. These coefficients are given by $\{c_0=1\}_{Flat}$, $\{c_0=0.5,c_1=0.5\}_{Hahn}$, and $\{c_0=0.42,c_1=0.5,c_2=0.08\}_{Blackman}$. Eqs.~\ref{Eq:lambda} and~\ref{Eq:F(f)} lead to

\begin{eqnarray}
\label{Eq:F(f0)}
F(f_0, \bar x)=\frac{1}{\Delta x}\sum_nc_n\sum_{\epsilon=\pm 1}
\int_{x_m}^{x_M} A_0(x)\sin(2\pi f_0x + \phi)\e^{-2i\pi f_0 x-2i\pi n\epsilon (x-\bar x)/\Delta x}dx
\end{eqnarray}

for $f = f_0$, $x_M = \bar x + \Delta x/2$, $x_m = \bar x - \Delta x/2$  and $A_0(x)=a_0\e^{-\lambda x}$. Writing
$F(f_0,\bar x)=\sum_nc_n\sum_{\epsilon=\pm 1}F_{n\epsilon}$ in Eq.~\ref{Eq:F(f0)}, we
compute individually $F_{n\epsilon}$ which leads after integration to

\begin{eqnarray}\label{Eq:Fne}
F_{n\epsilon}=
\frac{2a_0\e^{2i\pi n\epsilon \bar x/\Delta x+i\phi}}{i\Delta x}
\left [
\e^{-\lambda_{n\epsilon}\bar x}\frac{\sinh(\lambda_{n\epsilon}\Delta x/2)}{\lambda_{n\epsilon}}
-\e^{-\Lambda_{n\epsilon}\bar x}\frac{\sinh(\Lambda_{n\epsilon}\Delta x/2)}{\Lambda_{n\epsilon}}
\right ]
\end{eqnarray}

where we have defined $\lambda_{n\epsilon}=\lambda+2i\pi n\epsilon/\Delta x$ and
$\Lambda_{n\epsilon}=\lambda+4i\pi f_0+2i\pi n\epsilon/\Delta x$. This expression does not depend on $\phi$ up to a global sign, for the values $\phi=0,\pi$. Assuming $f_0\gg \lambda$,
only the first term in bracket will contribute to $F_{n\epsilon}$. Since $\sinh(\lambda_{n\epsilon}\Delta x/2)=(-1)^n\sinh(\lambda\Delta x/2)$, one obtains

\begin{eqnarray}
F_{n\epsilon}\simeq
\frac{2a_0\e^{-\lambda \bar x+i\phi}}{i\Delta x}(-1)^n
\frac{\sinh(\lambda\Delta x/2)}{\lambda+2i\pi n\epsilon/\Delta x}
\end{eqnarray}

After summing over $\epsilon$, the Fourier transform finally is equal to

\begin{eqnarray}
\label{Eq:F}
F(f_0, \bar x)\simeq -iA_0(\bar x)\e^{i\phi}
\frac{\sinh(\lambda\Delta x/2)}{\lambda \Delta x/2}
\sum_{n\ge 0}(-1)^nc_n
\frac{(\lambda \Delta x/2)^2}{
(\lambda \Delta x/2)^2+\pi^2n^2}
\end{eqnarray}

The exact formula is obtained by incorporating the contribution from the second term of Eq.~\ref{Eq:Fne}, involving  $\Lambda_{n\epsilon}$ which induces oscillations as function of $\bar x$ and $\Delta x$, with frequency $f_0$:

\begin{eqnarray}
\label{Eq:Ff0}
&F(f_0, \bar x)= -iA_0(\bar x)\e^{i\phi}
\sum_{n\ge 0}(-1)^nc_n
\Big [
\frac{\sinh(\lambda\Delta x/2)}{\lambda \Delta x/2}
\frac{(\lambda \Delta x/2)^2}{
(\lambda \Delta x/2)^2+\pi^2n^2}
\\ \nonumber
&-\e^{-4i\pi f_0\bar x}
\left [\cos(2\pi f_0\Delta x)\sinh(\lambda\Delta x/2)+i\sin(2\pi f_0\Delta x)\cosh(\lambda\Delta x/2)\right ]
\frac{(\lambda+4i\pi f_0) \Delta x/2}{
\{(\lambda+4i\pi f_0) \Delta x/2\}^2+\pi^2n^2}
\Big ]
\end{eqnarray}

As $\Delta x$ goes to zero for finite $\lambda$ in Eq.~\ref{Eq:F}, it yields the Fourier amplitude as $|F(f_0, \bar x)|\simeq A_F( \bar x) c_0 $. As a result, one defines $A_F( \bar x) = |F(f_0, \bar x)|/c_0$, in order to normalize the function with respect to $A_0(\bar x)$ in this limit. While $\sinh(\lambda \Delta x/2)/(\lambda  \Delta x/2)$ goes to 1 as $\Delta x$ goes to zero, the other contributions in Eq.~\ref{Eq:Ff0}, which involve oscillatory terms periodic in $\Delta x$ with the frequency $f_0$, grow simultaneously. They are responsible for the oscillatory behaviour reported in Figs.~\ref{Fig_ET4Zn_fenetre} and~\ref{Fig_f0}. In particular,
in this limit, if we take into account all the contributions in Eq.~\ref{Eq:Ff0}, one obtains

\begin{eqnarray}
\label{Eq:Ff00}
\lim_{\Delta x\rightarrow 0}F(f_0, \bar x)\simeq  2c_0A_0(\bar x)\e^{i\phi-2i\pi f_0\bar x}\sin(2\pi f_0\bar x)
\end{eqnarray}

Furthermore, as discussed in Section~\ref{sec:methodology}, $A_F( \bar x)/A_0( \bar x)$ only depends on $\lambda$ at a given $\Delta x$.


\begin{thebibliography}{0}%
\makeatletter
\providecommand \@ifxundefined [1]{%
 \@ifx{#1\undefined}
}%
\providecommand \@ifnum [1]{%
 \ifnum #1\expandafter \@firstoftwo
 \else \expandafter \@secondoftwo
 \fi
}%
\providecommand \@ifx [1]{%
 \ifx #1\expandafter \@firstoftwo
 \else \expandafter \@secondoftwo
 \fi
}%
\providecommand \natexlab [1]{#1}%
\providecommand \enquote  [1]{``#1''}%
\providecommand \bibnamefont  [1]{#1}%
\providecommand \bibfnamefont [1]{#1}%
\providecommand \citenamefont [1]{#1}%
\providecommand \href@noop [0]{\@secondoftwo}%
\providecommand \href [0]{\begingroup \@sanitize@url \@href}%
\providecommand \@href[1]{\@@startlink{#1}\@@href}%
\providecommand \@@href[1]{\endgroup#1\@@endlink}%
\providecommand \@sanitize@url [0]{\catcode `\\12\catcode `\$12\catcode
  `\&12\catcode `\#12\catcode `\^12\catcode `\_12\catcode `\%12\relax}%
\providecommand \@@startlink[1]{}%
\providecommand \@@endlink[0]{}%
\providecommand \url  [0]{\begingroup\@sanitize@url \@url }%
\providecommand \@url [1]{\endgroup\@href {#1}{\urlprefix }}%
\providecommand \urlprefix  [0]{URL }%
\providecommand \Eprint [0]{\href }%
\providecommand \doibase [0]{http://dx.doi.org/}%
\providecommand \selectlanguage [0]{\@gobble}%
\providecommand \bibinfo  [0]{\@secondoftwo}%
\providecommand \bibfield  [0]{\@secondoftwo}%
\providecommand \translation [1]{[#1]}%
\providecommand \BibitemOpen [0]{}%
\providecommand \bibitemStop [0]{}%
\providecommand \bibitemNoStop [0]{.\EOS\space}%
\providecommand \EOS [0]{\spacefactor3000\relax}%
\providecommand \BibitemShut  [1]{\csname bibitem#1\endcsname}%
\let\auto@bib@innerbib\@empty
\end{thebibliography}%


\begin{thebibliography}{00}

\bibitem{Sh84} D. Shoenberg, Magnetic Oscillations in Metals (Cambridge University Press, Cambridge, 1984). 

\bibitem{On12} Y. Onuki and R. Settai, Low Temp. Phys. 38, 89 (2012). 

\bibitem{Te14} T. Terashima, N. Kikugawa, A. Kiswandhi, E.-S. Choi, J. S. Brooks, S. Kasahara, T. Watashige, H. Ikeda, T. Shibauchi, Y. Matsuda, T. Wolf, A. E. B
B\"{o}hmer, F. Hardy, C. Meingast, H. v. L\"{o}hneysen, M.-T. Suzuki, R. Arita and S. Uji, Phys. Rev. B \textbf{90} 144517 (2014) 

\bibitem{W15} M. D. Watson, T. K. Kim, A. A. Haghighirad, N. R. Davies, A. McCollam, A. Narayanan, S. F. Blake, Y. L. Chen, S. Ghannadzadeh, A. J. Schofield,
M. Hoesch, C. Meingast, T. Wolf and A. I. Coldea, Phys. Rev. B \textbf{91} 155106 (2015). 

\bibitem{Wa15}  M. D. Watson, T. Yamashita, S. Kasahara, W. Knafo, M. Nardone, J. B\'{e}ard, F. Hardy, A. McCollam, A. Narayanan, S. F. Blake, T. Wolf,
A. A. Haghighirad, C. Meingast, A. J. Schofield, H. v. L\"{o}hneysen, Y. Matsuda, A. I. Coldea and T. Shibauchi, Phys. Rev. Lett. \textbf{115} 027006 (2015). 

\bibitem{Au15f} A. Audouard, F. Duc, L. Drigo, P. Toulemonde, S. Karlsson, P. Strobel and A. Sulpice, EPL \textbf{109} 27003 (2015). 

\bibitem{Uj08} S. Uji and J. S. Brooks, Springer Series Material Science Vol. 110 (Springer, 2008), p. 89. 

\bibitem{Au13} A. Audouard and J.-Y. Fortin, C. R. Physique \textbf{14} 15 (2013). 

\bibitem{Au15} A. Audouard, J.-Y. Fortin, D. Vignolles,  R. B. Lyubovskii, L. Drigo, G. V. Shilov, F. Duc, E. I. Zhilyaeva, R. N. Lyubovskaya and
E. Canadell J. Phys.: Condens. Matter \textbf{27} 315601 (2015). 

\bibitem{Pi62} A. B. Pippard, Proc. Roy. Soc. (London) \textbf{A270} 1 (1962). 

\bibitem{Au12} A. Audouard, J.-Y. Fortin, D. Vignolles, R. B. Lyubovskii, L. Drigo, F. Duc, G. V. Shilov, G. Ballon, E. I. Zhilyaeva, R. N. Lyubovskaya and
E. Canadell, EPL \textbf{97} 57003 (2012). 

\bibitem{footnote1} At the lowest temperature (2 K) and highest field (55 T), i.e. lowest $u_0Tm_{\beta}x$ values explored in Ref.~\cite{Au15}, this approximation overestimates the amplitude of the considered $\beta$ component by less than 3\%. Accordingly, the relevant reported mass plots are linear (see also Fig.~\ref{Fig_ET4Zn_masseeffective}(a)).

\bibitem{Fi64} A.S. Filler, J. Opt. Soc. Am. \textbf{54}, 762 (1964). 

\bibitem{Al77} J. Allen , IEEE Trans. Acoust., Speech, Signal Processing, \textbf{ASSP-25} 235 (1977). 

\bibitem{Ha78} F.J. Harris, Proc. IEEE \textbf{66}, 51 (1978). 

\bibitem{Ac90} F. S. Acton, Numerical Methods that work, The Mathematical Association of America (Washington D.C., 1990). 

\bibitem{Na07} D.A. Naylor and M.K. Tahic, J. Opt. Soc. Am. A \textbf{24}, 3644 (2007). 

\bibitem{Dr10} L. Drigo, F. Durantel, A. Audouard and G. Ballon, Eur. Phys. J.-Appl. Phys. \textbf{52} 10401 (2010). 











\end{thebibliography}
 \end{document}